\documentclass[12pt]{iopart}
\usepackage{graphicx}

\begin{document}

\title{Detecting the overlapping and hierarchical community structure in complex networks}

\author{Andrea Lancichinetti}

\address{Complex Networks Lagrange Laboratory (CNLL),
Institute for Scientific Interchange (ISI), Viale S. Severo 65, 10133, Torino, Italy}
\ead{arg.lanci@gmail.com}

\author{Santo Fortunato}

\address{Complex Networks Lagrange Laboratory (CNLL),
Institute for Scientific Interchange (ISI), Viale S. Severo 65, 10133, Torino, Italy}
\ead{fortunato@isi.it}

\author{J\'anos Kert\'esz}

\address{Department of Theoretical Physics, Budapest University of Technology and Economics, Budafoki \'ut 8, H1111,
Budapest, Hungary}
\ead{kertesz@phy.bme.hu}

\begin{abstract}
Many networks in nature, society and technology are characterized by a 
mesoscopic level of organization, with groups of nodes forming tightly connected units, 
called communities or modules, that are only weakly linked to each other. Uncovering this 
community structure is one of the most important problems in the field of complex networks. 
Networks often show a hierarchical organization, with communities embedded within 
other communities; moreover, nodes can be shared between different communities. Here we 
present the first algorithm that finds both overlapping communities and the hierarchical 
structure. The method is based on the local optimization of a fitness function.
Community structure is revealed by peaks in the fitness histogram. 
The resolution can be tuned by a parameter enabling to investigate different hierarchical levels of organization.
Tests on real and artificial networks give excellent results. 

\end{abstract}
\maketitle

\section{Introduction}

The study of networks as the "scaffold of complexity" has proved very successful to understand 
both the structure and the function of many natural and artificial  
systems~\cite{bara02,mendes03,Newman:2003,psvbook,vitorep}. A common feature of complex networks 
is {\it community structure}~\cite{Girvan:2002, Newman:2004,arenasrev,miareview}, i.e., 
the existence of groups of nodes such that nodes within a group are much more connected to each 
other than to the rest of the network. Modules or communities reflect topological relationships 
between elements of the underlying system and represent functional entities. E.g., communities 
may be groups of related individuals in social networks~\cite{Girvan:2002, Lusseau:2005, Adamic:2005}, 
sets of Web pages dealing with the same topic~\cite{Flake:2002}, taxonomic categories in food 
webs~\cite{foodw1,foodw2}, biochemical pathways in metabolic networks~\cite{Holme:2003,Guimera:2005,palla}, 
etc. Therefore the identification of communities is of central importance but it has remained a formidable task.

The solution is hampered by the fact that the organization of networks at the "mesoscopic", 
modular level is usually highly non-trivial, for at least two reasons. First, there is often a whole 
hierarchy of modules, because communities are nested: Small communities build larger ones, 
which in turn group together to form even larger ones, etc. 
An example could be the organization of 
a large firm, and it has been argued that the complexity of life can also be mapped to a hierarchy 
of networks \cite{pyramid}. The hierarchical form of organization can be very efficient, with the 
modules taking care of specific functions of the system~\cite{clauset07}. In the presence of hierarchy, 
the concept of community structure becomes richer, and demands a method that is able to detect all 
modular levels, not just a single one. Hierarchical clustering is a well-known technique 
in social network analysis~\cite{wasserman,scott}, biology~\cite{biol} and finance~\cite{mantegna}.
Starting from a partition in 
which each node is its own community, or all nodes are in the same community, one merges or 
splits clusters according to a topological measure of similarity between nodes. In this way one 
builds a hierarchical tree of partitions, called {\it dendrogram}. Though this method naturally 
produces a hierarchy of partitions, nothing is known {\it a priori} about their qualities.
The modularity of Newman and Girvan~\cite{Newman:2004b} is a measure of the quality of a partition, but 
it can identify a single partition, i.e. the one corresponding to the largest value of the measure. 
Only recently scholars have
started to focus on the problem of identifying meaningful hierarchies~\cite{clauset07,sales07}.

\begin{figure}[htb]
\begin{center}
\includegraphics[width=8cm]{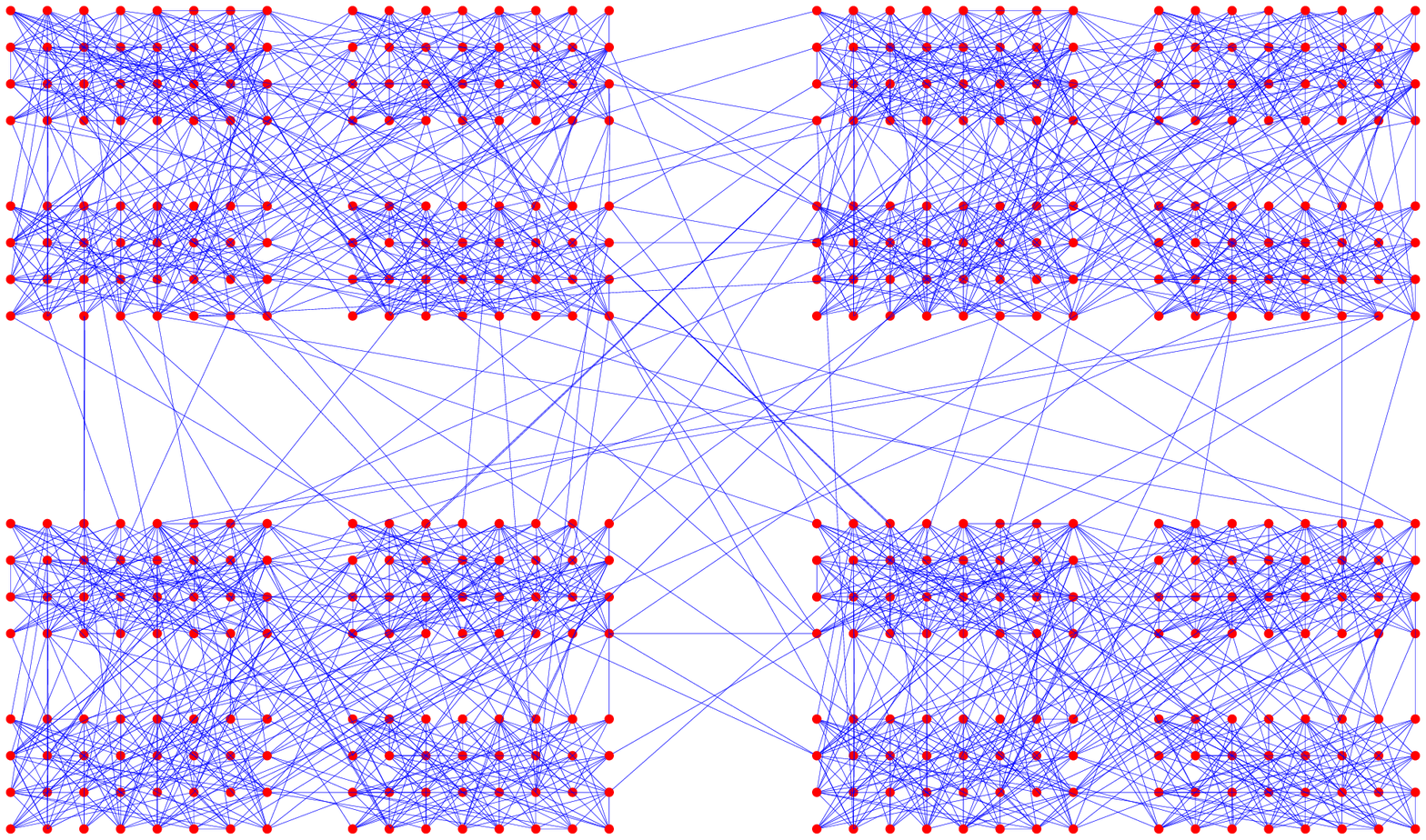}\\
\vspace{1cm}
\includegraphics[width=4cm]{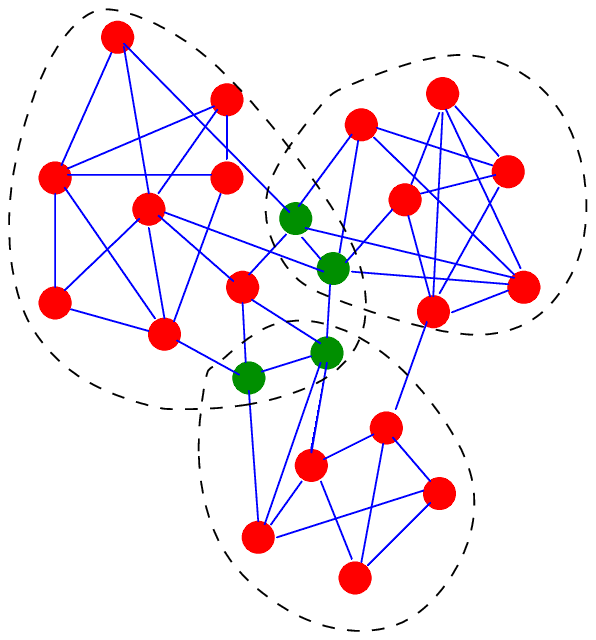}
\caption{\label{fig1} (Top) A network with a hierarchical structure. Each of the four large clusters is made out
of $128$ nodes and has an internal subdivision in four clusters with $32$ nodes. (Bottom) 
Overlapping communities. The green nodes are topologically related to more groups.}
\end{center}
\end{figure}

A second difficulty is caused by the fact that nodes often 
belong to more than one module, 
resulting in overlapping communities~\cite{palla,baumes1,baumes2,zhang,nicosia}. For instance people 
belong to different social communities, depending on their families, friends, professions, hobbies, 
etc. Nodes belonging to more than one community are a problem for standard methods and 
lower the quality of the detected modules. Moreover, this conceals important information, 
and often leads to misclassifications. The problem of overlapping communities was exposed 
in~\cite{palla}, where a solution to it was also offered. The proposed method is based on clique 
percolation: A $k$-clique (a complete subgraph of $k$ nodes) is rolled over the network by using 
$k-1$ common nodes. This way a set of nodes can be reached, which is identified as a community. 
One node can participate in more than one such unit, therefore overlaps naturally occur. 
The method however is not suitable for the detection of hierarchical structure, as
once the size $k$ of the clique is specified, mostly a single modular structure can be recovered.
In Fig.~\ref{fig1} we show distinct networks with hierarchical structure and overlapping communities, 
though it should be emphasized that these features often occur simultaneously. 

In order to provide the most exhaustive information about the modular structure of a graph,
a good algorithm should be able to detect both overlapping communities and hierarchies between them.
In this paper we introduce a framework that accomplishes these two demands. The method performs a local exploration
of the network, searching for the natural community of each node. During the procedure, 
nodes can be visited many times, no matter whether they have been assigned to a community or not.
In this way, overlapping communities are naturally recovered. The variation of a resolution parameter, 
determining the average size of the communities, allows to explore all hierarchical levels
of the network.
\vskip0.4cm

\section{The method}
\label{sec1}

The basic assumption behind our algorithm is that communities are essentially local structures, 
involving the nodes belonging to the modules themselves plus at most an extended neighborhood of 
them. This is certainly plausible for large networks, where each node does not depend on most of 
its peers. In the link graph of the WWW, for instance, one does not even have a perception of how 
large the network is, and topical communities are formed based only on partial information about 
the graph. Similarly, social communities are local structures without any reference to the humankind as a whole.

Here a community is a subgraph identified by the maximization of a property or 
{\it fitness} of its nodes. We have tried several options for the 
form of the fitness and obtained the best results with the simple expression
\begin{equation}
f_{\cal G}=\frac{k_{in}^{\cal G}}{(k_{in}^{\cal G}+k_{out}^{\cal G})^\alpha},
\label{eq1}
\end{equation}
where $k_{in}^{\cal G}$ and $k_{out}^{\cal G}$ are the total internal and external degrees of the nodes 
of module $\cal G$, and $\alpha$ is a positive real-valued parameter, controlling the size 
of the communities. The internal degree of a module equals the double of the number of 
internal links of the module. The external degree is the number of links joining each member 
of the module with the rest of the graph. The aim is to determine a 
subgraph starting from node $A$ such that the inclusion of a new node, or the elimination 
of one node from the subgraph would lower $f_{\cal G}$. We call such subgraph the {\it natural community} of node $A$.
This amounts to determine 
local maxima for the fitness function for a given $\alpha$. The true maximum for each node 
trivially corresponds to the whole network, because in this case $k_{out}^{\cal G}=0$ and 
the value of $f_{\cal G}$ is the largest that the measure can possibly attain for a given 
$\alpha$. The idea of detecting communities by a local optimization of some metric has already 
been applied earlier~\cite{baumes1,baumes2,clauset05,bagrow05}.

It is helpful to introduce the concept of node fitness. Given a fitness function, the fitness 
of a node $A$ with respect to subgraph $\cal G$, $f_{\cal G}^A$, is defined as the variation 
of the fitness of subgraph $\cal G$ with and without node $A$, i.e.
\begin{equation}
f_{\cal G}^A=f_{{\cal G}+\{A\}}-f_{{\cal G}-\{A\}}.
\label{eq2}
\end{equation}
In Eq.~\ref{eq2}, the symbol ${\cal G}+\{A\}$ (${\cal G}-\{A\}$) indicates the subgraph 
obtained from module $\cal G$ with node $A$ inside (outside). 

\begin{figure}[htb]
\begin{center}
\includegraphics[width=7cm]{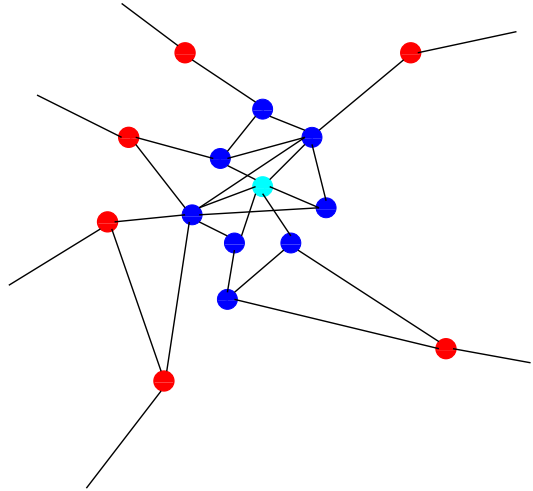}
\caption{\label{fig2} Schematic example of natural community for a node (sky-blue point in the figure) 
according to our definition. The blue nodes are the other members of the group and 
have positive fitness within the group, while the red nodes have all negative fitness with respect to the group.}
\end{center}
\end{figure}

The natural community of a node $A$ is identified with the following procedure. Let us suppose that we
have covered a subgraph $\cal G$ including node $A$. Initially, $\cal G$ is identified with 
node $A$ ($k_{in}^{\cal G}=0$). Each iteration of the algorithm consists of the following steps:
\begin{enumerate}
\item{a loop is performed over all neighboring nodes of $\cal G$ not included in $\cal G$;}
\item{the neighbor with the largest fitness is added to $\cal G$, yielding a larger subgraph ${\cal G}^\prime$;}
\item{the fitness of each node of ${\cal G}^\prime$ is recalculated;}
\item{if a node turns out to have negative fitness, it is removed from ${\cal G}^\prime$, yielding a new subgraph
${\cal G}^{\prime\prime}$;}
\item{if 4 occurs, repeat from 3, otherwise repeat from 1 for subgraph ${\cal G}^{\prime\prime}$.}
\end{enumerate}
The process stops when the nodes examined in step $1$ all have negative fitness (Fig.~\ref{fig2}). 
This procedure corresponds
to a sort of greedy optimization of the fitness function, as at each move one looks for the highest possible increase. Other recipes
may be adopted as well. For instance one could backtrack nodes with negative fitness only when the cluster stops growing and/or
include in the cluster the first neighboring node that produces an increase fo the fitness. Such recipes may lead to 
higher fitness clusters in a shorter time, and deserve in-depth investigations, which we leave for future work.

We define a {\it cover}
of the graph as a set of clusters such that each node is assigned to {\it at least} one cluster. 
This is an extension of the traditional concept of graph partition (in which each node
belongs to a single cluster), to account for possible overlapping communities. 
In our case, detecting a cover amounts to
discovering the natural community of each node of the graph at study. 
The straightforward way to achieve this is to repeat the above procedure for each single node. 
This is, however, computationally expensive. The natural communities of many nodes often coincide, 
so most of the computer time is spent to rediscover the same modules over and over. An economic 
way out can be summarized as follows:
\begin{enumerate}
\item{pick a node $A$ at random;}
\item{detect the natural community of node $A$;}
\item{pick at random a node $B$ not yet assigned to any group;}
\item{detect the natural community of $B$, exploring all nodes regardless of their possible
membership to other groups;} 
\item{repeat from $3$.} 
\end{enumerate}
The algorithm stops when all nodes have been assigned to at least one group.
Our recipe is justified by the following argument. The nodes of every community are either
overlapping with other communities or not. The community was explored about a specific node; if one 
chose any of the other nodes one would either recover the same community or one of the possible overlapping
communities. But the latter can be found as well if one starts from nodes which are outside the community at hand
and non-overlapping with it. In this way one should recover all modules, without needing to start from every node.
At the same time, overlapping nodes will be covered during the construction of each 
community they belong to, as it is possible to include nodes already assigned to other modules. 
Extensive numerical tests show that the loss in accuracy is minimal if one proceeds as we suggest rather than 
by finding the natural community of all nodes. We remark that the procedure has some degree 
of stochasticity, due to the random choice of the node-seeds from which communities are closed. 
The issue is discussed in Appendix~\ref{app1}.

The parameter $\alpha$ tunes the resolution of the method. Fixing $\alpha$ means setting 
the scale at which we are looking at the network. Large values of $\alpha$ yield very 
small communities, small values  
instead deliver large modules. If $\alpha$ is small enough, all nodes end up in the same 
cluster, the network itself. We have found that, in most cases, for $\alpha<0.5$ there is only one community,
for $\alpha>2$ one recovers the smallest communities.  
A natural choice is $\alpha=1$, as it is the ratio of the external degree to the total degree of the 
community. This corresponds to the so-called weak definition of community 
introduced by Radicchi et al.~\cite{radicchi}. We found that in most cases the cover found for $\alpha=1$
is relevant, so it gives useful information about the actual community structure of the graph at hand.
Sticking to a specific value of $\alpha$ means constraining the resolution of the method, much like it happens by
optimizing Newman-Girvan's modularity~\cite{FB,Kumpula}. 
However, one cannot know {\it a priori} how large the communities are, as this is one of the unknowns of 
the problem, so it is necessary to compare covers obtained at different scales.

By varying 
the resolution parameter one explores the whole hierarchy of covers of the graph, 
from the entire network down to the single nodes, leading to the most complete information 
on the community structure of the network. However, the method gives as well a natural way 
to rank covers based on their relevance. It is reasonable to think that a 
good cover of the network is {\it stable}, i.e. can be destroyed only by 
changing appreciably the value of $\alpha$ for which it was recovered. Each cover 
is delivered for $\alpha$ lying within some range. A stable cover would be 
indicated by a large range of $\alpha$. What we need is a quantitative index to label a 
cover $\cal P$. We shall adopt the average value $\bar{f}_{\cal P}$ of the fitness of its communities, i.e.
\begin{equation}
\bar{f}_{\cal P}=\frac{1}{n_c}\sum_{i=1}^{n_c}f_{{\cal G}_i}(\alpha=1),
\label{eq3}
\end{equation}
where $n_c$ is again the number of modules. The fitness must be calculated for a fixed 
value of $\alpha$ (we choose $\alpha=1$ for simplicity), such that identical (different) 
covers can be recognized by equal (different) values.
We shall derive the histogram
of the $\bar{f}_{\cal P}$-values of the covers 
obtained for different $\alpha$-values: stable covers are revealed by pronounced peaks 
in the resulting fitness histogram. The higher the peak, the more stable the cover. In this way,
covers can be ranked based on their frequency. A similar
concept of stability has been adopted in a recent study where a resolution parameter 
was introduced in Newman-Girvan's modularity~\cite{arenas08}.

A natural question is how to combine hierarchical communities with overlapping communities, as the usual
meaning of hierarchies seems incompatible with the existence of nodes shared among communities. However, this is only apparent
and the same definition of hierarchical partitions can be extended to the case of overlapping communities. We say that two partitions
$\mathcal{C'}$ and $\mathcal{C''}$ are {\it hierarchically ordered}, with $\mathcal{C'}$ higher than $\mathcal{C''}$,
if all nodes of each community of $\mathcal{C''}$ participate (fully or partially) in a single community of partition $\mathcal{C'}$.

\begin{figure}[htb]
\begin{center}
\includegraphics[width=\columnwidth]{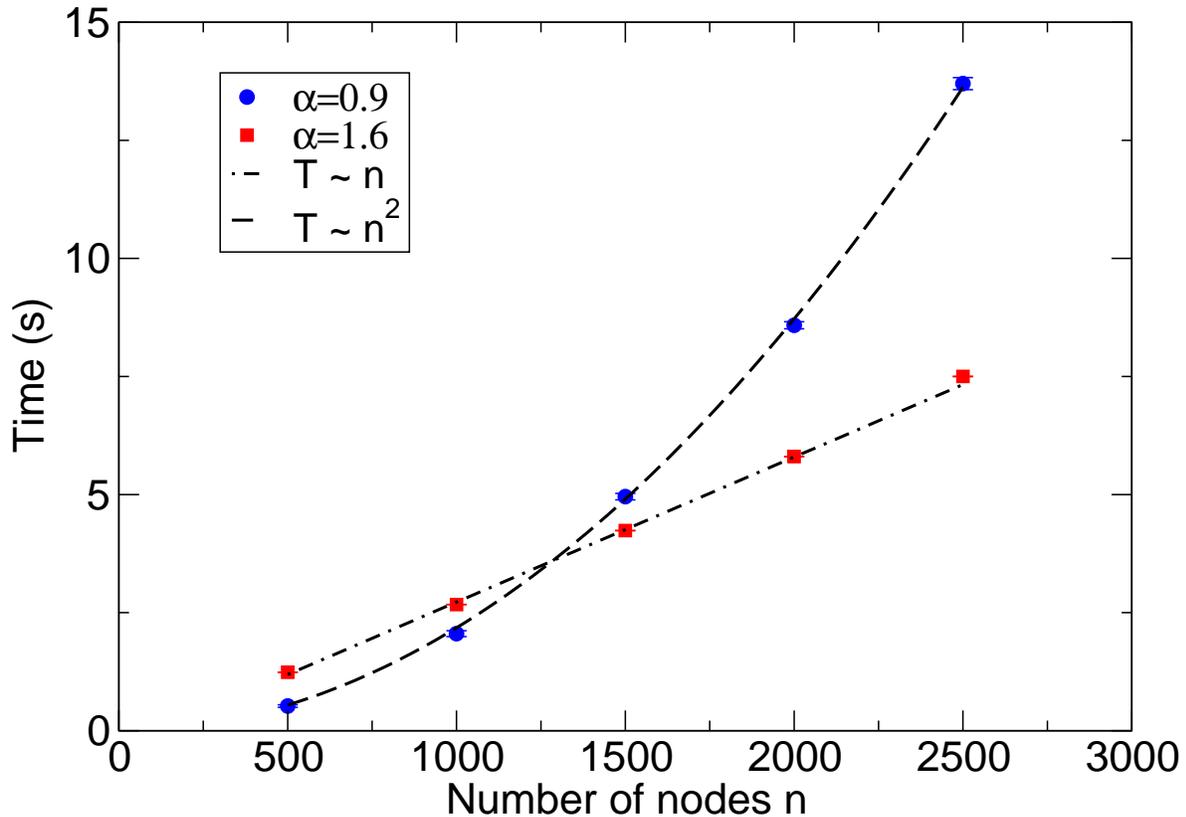}
\caption{\label{fig2a}Computational complexity of the algorithm. The two curves show how the time
to run the algorithm scales with the size of the graph for Erd\"os-R\'enyi networks with average degree $10$, for
$\alpha=0.9$ and $1.6$, respectively. The complexity ranges from quadratic for $\alpha=0.9$, for which communities are sizeable,
to linear for $\alpha=1.6$, for which communities are small.}
\end{center}
\end{figure}

It is hard to estimate the computational complexity of the algorithm, as it depends 
on the size of the communities and the extent of their overlaps, which in turn strongly 
depend on the specific network at study along with the value of the parameter $\alpha$. 
The time to build a community with $s$ nodes scales approximately as $O(s^2)$, due to the backtracking steps.
Therefore, a rough estimate of the complexity for a fixed $\alpha$-value is $O(n_c<s^2>)$, where $n_c$ is the number of modules 
of the delivered cover and $\langle s^2\rangle$ the second moment of the community size. 
The square comes from the loop over all nodes of a community to check for their fitness 
after each move. The worst-case complexity is $O(n^2)$, where $n$ is the number of nodes 
of the network, when communities are of size comparable with $n$. This is in general not 
the case, so in most applications the algorithm runs much faster and almost linearly when communities are small. 
The situation is shown in Fig.~\ref{fig2a}, where we plot the time to run the algorithm to completion for two 
different $\alpha$-values as a function of the number of nodes for Erd\"os-R\'enyi graphs with average degree $10$: the complexity
goes from quadratic to linear.
The total complexity of the algorithm to perform the complete analysis of a network depends as well on the number of 
$\alpha$-values required to resolve its hierarchical structure.
The hierarchy of covers can be the better displayed, the larger the number of $\alpha$-values 
used to run the algorithm. 
If the network has a hierarchical structure, 
as it often happens in real systems, the number of covers grows as $\log n$. 
In this case, the number of different $\alpha$-values required to resolve the hierarchy 
is also of the order of $\log n$ and the complete analysis can be carried out very quickly. 
We note that each iteration of the algorithm for a given $\alpha$ is independent of the others.
So, the calculation can be trivially parallelized by running different $\alpha$-values on 
each computer. If large computer resources are not available, a cheap way to proceed could 
be to start from a large $\alpha$-value, for which the algorithm runs to completion in a 
very short time, and use the final cover as initial configuration for a run at 
a slightly lower $\alpha$-value. Since the corresponding cover is similar to 
the initial one, also the second run would be completed in a short time and one can repeat 
the procedure all the way to the left of the range of $\alpha$. 

We conclude that for hierarchical networks our procedure has a 
worst-case computational complexity of $n^2\log n$. We remark that, if it is true that several algorithms nowadays
have a lower complexity, none of them is capable to carry out a complete analysis of the hierarchical community structure,
as they usually deliver a single partition. Therefore a fair comparison is not possible. Besides, other recipes for the local
optimization of our or other fitness functions may considerably lower the computational complexity of the algorithm, which seems a promising
research direction for the future.

\section{Results}
\label{sec2}

We extensively tested our method on artificial networks with built-in hierarchical 
community structure. We adopted a benchmark similar to that recently proposed by 
Arenas et al.~\cite{arenas06,arenas07}, which is a simple extension of the classical 
benchmark proposed by Girvan and Newman~\cite{Girvan:2002}. There are $512$ nodes, 
arranged in $16$ groups of $32$ nodes each. The 16 groups are ordered into 4 supergroups.
Every node has an average of $k_1$ links with the $31$ partners of its group and $k_2$ 
links with the $96$ nodes of other three groups within the same supergroup. In addition, 
each node has a number $k_3$ of links with the rest of the network. In this way, two 
hierarchical levels emerge: one consisting of the $16$ small groups, and one of the 
supergroups with $128$ nodes each (Fig.~\ref{fig1}top is an example). 
The degree of mixing of the four supergroups is 
expressed by the parameter $k_3$, that we tune freely. In principle we could also tune 
the mixing of the small communities, by varying the ratio $k_1/k_2$, but we prefer to 
set $k_1=k_2=16$, so that the micro-communities are ``fuzzy'', i.e. very mixed with each other, 
and pose a hard test to our method.
\begin{figure}[htb]
\begin{center}
\includegraphics[width=\columnwidth]{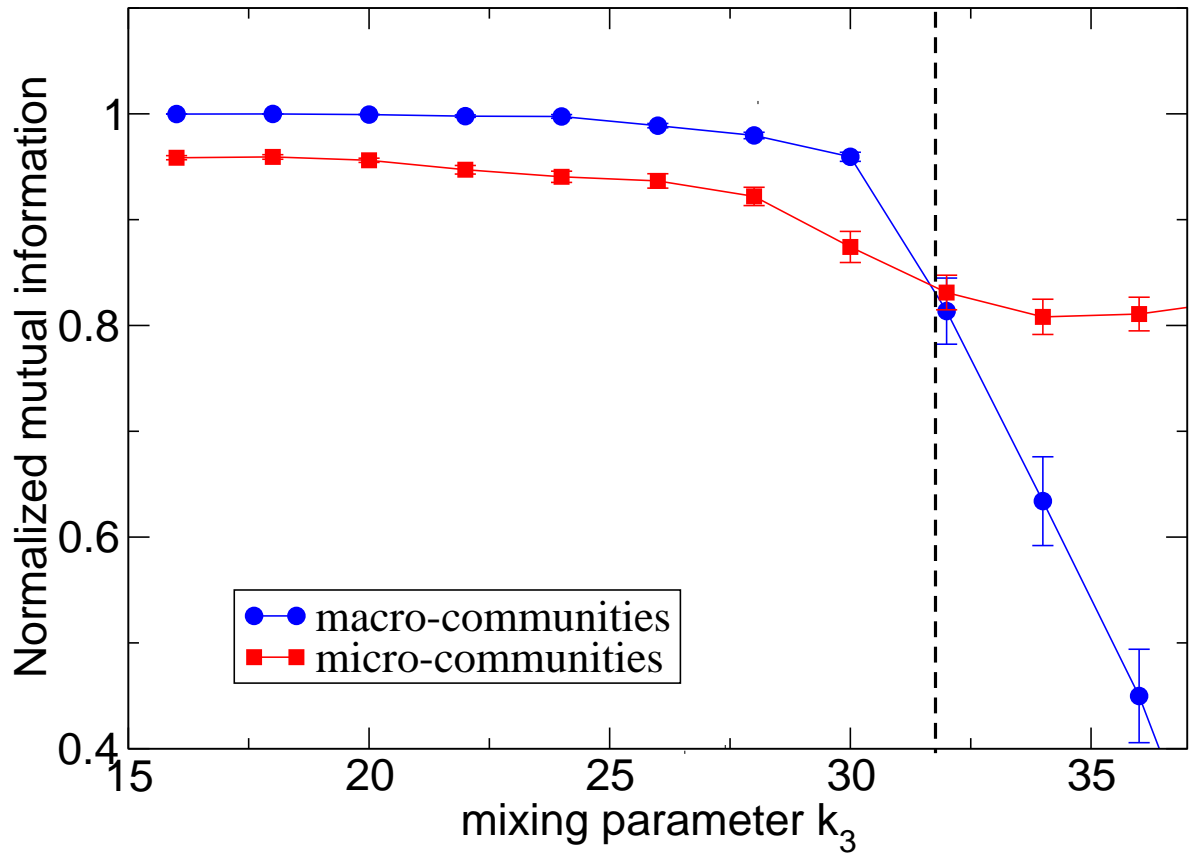}
\caption{\label{fig3} Test of the accuracy of our method on a hierarchical benchmark.  
The normalized mutual information is used to compare the cover
found by the algorithm with the natural cover of the network at each level. 
At the higher level (circles), the communities are four clusters including each four 
clusters of $32$ nodes, for a total of $128$ nodes per cluster. Our method finds
the right clusters as long as they are not too mixed with each other.
At the lower level (squares), the 
communities are $16$ clusters of $32$ nodes each. The method performs very well, considering that
each node has as many links inside as outside each micro-community, for any value of $k_3$. The dashed
vertical line marks the graph configurations for which the number of links of each node within its macro-community
equals the number of links connecting the node to the other three macro-communities.}
\end{center}
\end{figure}

We have to check whether the built-in hierarchy is recovered through the algorithm. 
This in general depends on the parameter $k_3$. Therefore, we considered different 
values of $k_3$: for each value we built $100$ realizations of the network. To compare 
the built-in modular structure with the one delivered by the algorithm
we adopt the {\it normalized mutual information}, a measure of 
similarity borrowed from information theory, which has proved to be 
reliable~\cite{Danon:2005}. The extension of the measure to overlapping communities is not trivial
and there are several alternatives. Our extension is discussed in Appendix~\ref{app2}. 
In Fig.~\ref{fig3} we plot the average value of the normalized mutual information as a function of $k_3$ 
for the two hierarchical levels. We see that in both cases the results are very good. 
The cover in the four supergroups or macro-communities is correctly identified 
for $k_3<24$, with very few exceptions, and the algorithm starts to fail only when 
$k_3\sim 32$, i.e., when each node has $32$ links inside and $~32$ outside of its 
macro-community, which is then very mixed with the others. The performance is very 
good as well for the lower hierarchical level: The modules are always well mixed 
with each other, as $k_1=k_2=16$ for any value of $k_3$, so it is remarkable that 
the resulting modular structure found by the algorithm is still so close to the 
built-in modular structure, up until $k_3\sim 32$. 
\begin{figure}[htb]
\begin{center}
\includegraphics[width=\columnwidth]{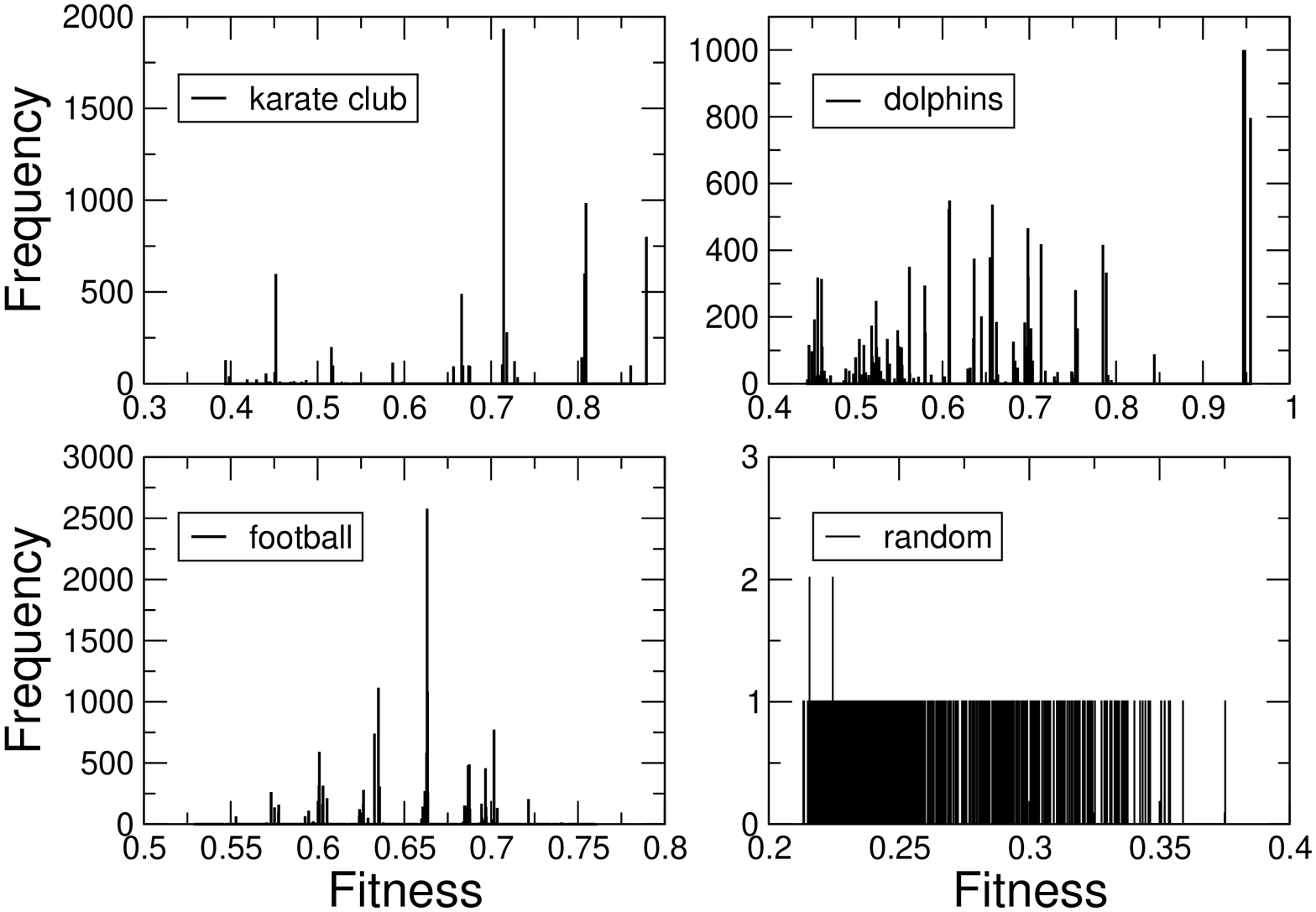}
\caption{\label{fig4a} Analysis of real networks. The fitness histograms correspond to Zachary's karate club (top-left),
Lusseau's dolphins' network (top-right) and the network of American college football teams (bottom-left). 
The highest peaks indicate the best covers, which coincide with
the natural covers of the graphs, except for Zachary's karate club, where it corresponds to the same split in four clusters 
found through modularity optimization. However, the social cover in two of the network is the third most relevant
cover. In (bottom-right) we show the fitness histogram for an Erd\"os-R\'enyi random graph with $100$ nodes and the same
average degree of the network of American college football teams: there is no visible
structure, as expected.}
\end{center}
\end{figure}
The main problem with this type of tests is that one does not have independent 
information about the covers, therefore it can be judged only if they 
are reasonable or not. Fortunately, for a few networks, covers have 
been identified by special information on the system itself and/or its history. 
In Fig.~\ref{fig4a} we show the fitness histograms corresponding to some of these networks, often used
to test algorithms: Zachary's karate club~\cite{zachary} (top-left), Lusseau's dolphins' network~\cite{Lusseau:2003} (top-right)
and the network of American college football teams~\cite{Girvan:2002} (bottom-left). 
The social network of karate club members studied by the 
sociologist Wayne Zachary has become a benchmark 
for all methods of community detection. The network consists of $34$ nodes, 
which separated in two distinct groups due to a contrast between one 
of the instructors and the administrator of the club.
The question is whether one is able to detect the social fission of the 
network. The second network represents
the social interactions of bottlenose dolphins living in Doubtful Sound, New Zealand. The network was
studied by the biologist David Lusseau, who divided the dolphins in two groups
according to their age. The third example is the network of American 
college football teams. Here, there are $115$ nodes, representing the 
teams, and two nodes are connected if their teams play against each other. 
The teams are divided into $12$ conferences. Games between teams in the same 
conference are more frequent than games between teams of different conferences, 
so one has a natural cover where the communities correspond to the 
conferences.

\begin{figure}[htb]
\begin{center}
\includegraphics[width=6.5cm]{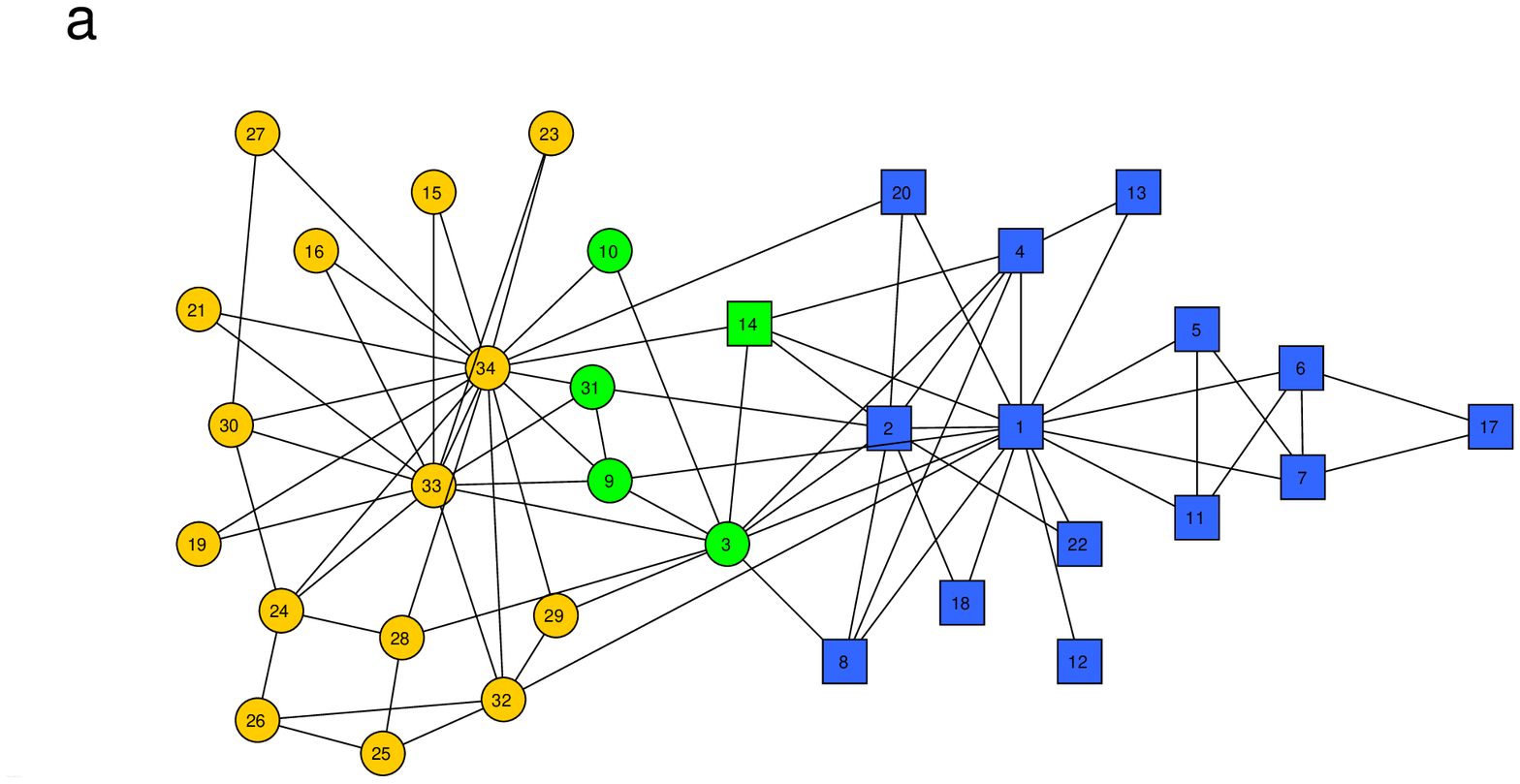}\\
\vspace{0.7cm}
\includegraphics[width=6.5cm]{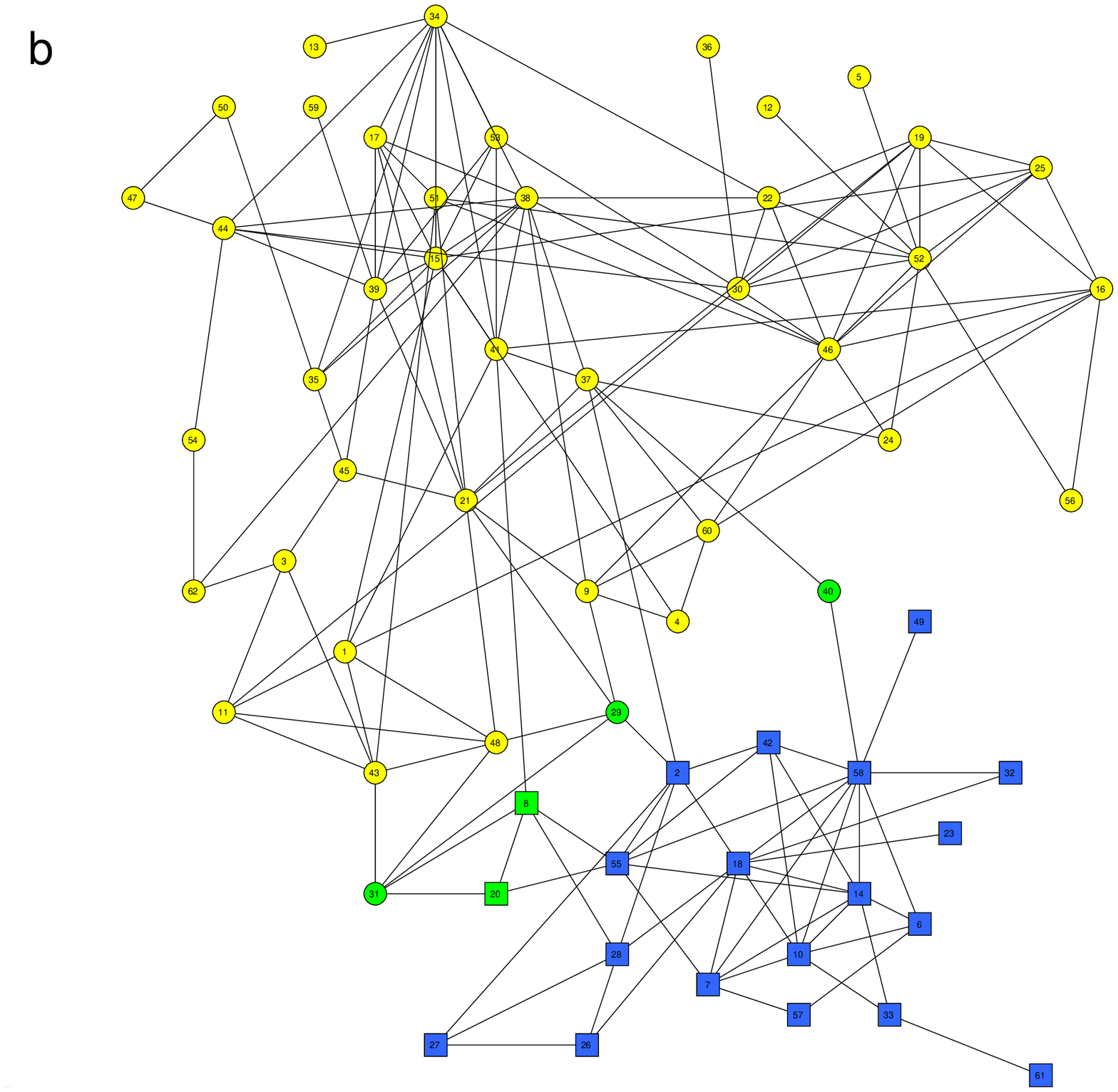}
\caption{\label{fig4} (a) Zachary's karate club. We show the hierarchical
levels corresponding to the cover in two clusters ($0.76<\alpha<0.84$). The nodes
$3$, $9$, $10$, $14$ and $31$ are shared between the two groups: 
such nodes are often misclassified by traditional algorithms. The non-overlapping nodes reflect the
social fission observed by Zachary, which is illustrated by the squares and the circles in the figure. 
(b) Lusseau's network of bottlenose dolphins. The best cover in two clusters that we found ($0.77<\alpha<0.82$)
agrees with the separation observed by Lusseau (squares and circles in the figure). The nodes
$8$, $20$, $29$, $31$ and $40$ are shared between the two groups.}
\end{center}
\end{figure}

The pronounced spikes in the histograms of Fig.~\ref{fig4a} 
show that these networks indeed have community structure. 
For Zachary's network we find that 
the most stable cover consists of four clusters. Even if this is 
not what one would like to recover, we stress that this cover is often found 
by other methods, like modularity optimization, which indicates that it is 
topologically meaningful. But our method can do better: The social split in 
two clusters (Fig.~\ref{fig4}a) turns out to be a higher hierarchical level, given by a pairwise 
merging of the four communities of the main cover. Interestingly, 
we found that the two groups are overlapping, sharing a few nodes. 
For the dolphins' network the highest spike corresponds to Lusseau's subdivision of the animals' population in two communities, with some
overlap between the two groups (Fig.~\ref{fig4}b). Similarly, the highest spike in Fig.~\ref{fig4a} (bottom-left) corresponds to the 
natural partition of the teams in conferences.

\begin{figure}[htb]
\begin{center}
\includegraphics[width=\columnwidth]{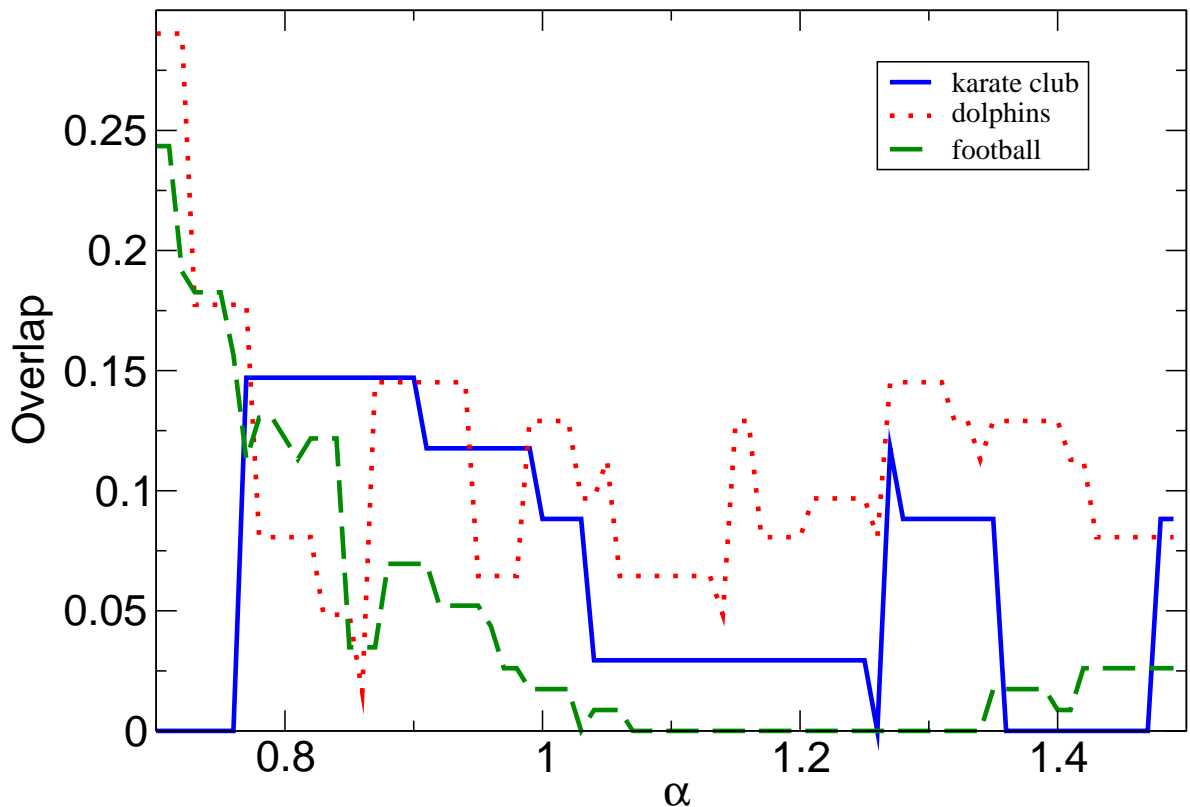}
\caption{\label{fig5a} Fraction of overlapping nodes as a function of $\alpha$ for the three real networks discussed
in Fig.~\ref{fig4a}: Zachary's karate club and the networks of dolphins and American football teams. 
There is no unique pattern, the extent of the overlap does not show a systematic 
variation with $\alpha$.}
\end{center}
\end{figure}

To check how good our algorithm performs as compared to other methods we have analyzed the karate, dolphins and American college football
networks with the clique percolation method (CPM) by Palla et al.~\cite{palla}. The values of the normalized mutual information of 
the covers found by the algorithm with respect to the real covers are 0.690 (our method) and 0.170 (CPM)
for the karate club, 0.781 (our method) and 0.254 (CPM) for the dolphins' network, 
0.754 (our method) and 0.697 (CPM) for the American college football network.
So our method proves superior to the CPM in these instances. On the other hand, the CPM performs better for 
networks with many cliques. An example is represented by the word association network built on the University of South Florida Free
Association Norms~\cite{nelson98}, analyzed in~\cite{palla}. The CPM finds groups of words which correspond to well defined categories,
whereas with our method the categories are more mixed. An important reason for this discrepancy is that our method recovers
overlapping nodes that usually lie at the border between communities, whereas in the word association network
they often are central nodes of a community. For instance, the word ``color'' is the central hub of the community
of colors, but it also belongs to other categories like ``Astronomy'' and ``Light''.

We performed tests on many other real systems, including protein interaction networks, 
scientific collaboration networks, and other social networks. In all cases we found reasonable covers.
On the other hand, we found that random graphs have no natural community structure, as covers
are unstable (Fig.~\ref{fig4a}, bottom-right). This is remarkable, as it is known that 
other approaches may find covers in random graphs as well~\cite{Guimera:2004}, 
a problem that triggered an ongoing debate 
as to when a cover is indeed relevant~\cite{karrer08}.

In Fig.~\ref{fig5a} we study how the extent of the overlap between the communities depends on the
resolution parameter $\alpha$, for three real networks. From the figure it is not possible to
infer any systematic dependence of the overlap on $\alpha$, the pattern is strongly dependent on the specific graph topology.

\begin{figure}[htb]
\begin{center}
\includegraphics[width=\columnwidth]{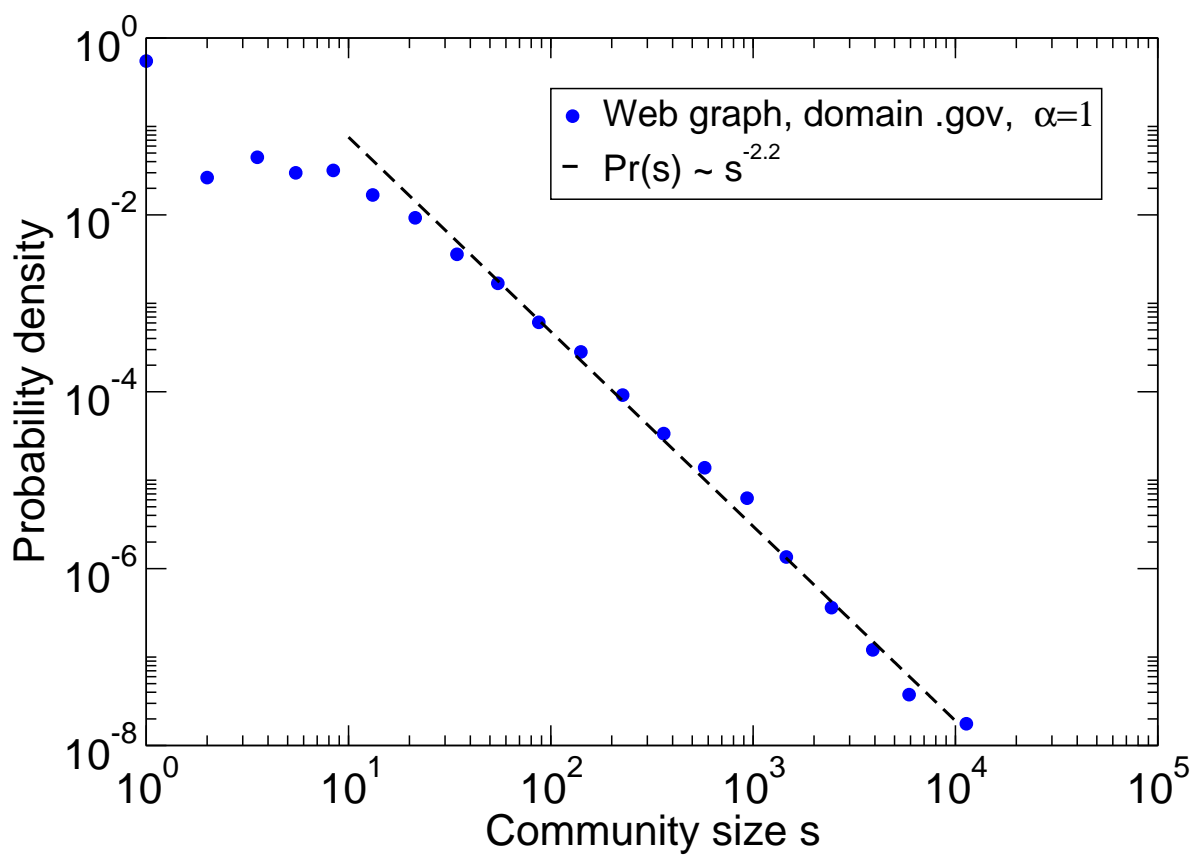}
\caption{\label{fig5} Distribution of community sizes for the link graph corresponding to the 
domain {\tt .gov} of the WWW. The resolution parameter $\alpha=1$. The distribution is clearly skewed,
in agreement with previous findings on large graphs. The tail can be well fitted by a power law with 
exponent $2.2(1)$ (dashed line in the figure).}
\end{center}
\end{figure}

We conclude the section with an analysis of the statistical properties of community structure in graphs.
Fig.~\ref{fig5} shows the distribution of community sizes for a sample of the WWW link graph, corresponding
to the subset of Web pages within the domain {\tt .gov}. We analyzed the largest connected component of 
the graph, consisting of $774,908$ nodes and $4,711,340$ links. 
The figure refers to the cover found for $\alpha=1$, which was 
identified within less than $40$ hours of CPU time on a small PC. 
The distribution of community sizes is skewed, with a tail that follows a power 
law with exponent $2.2(1)$. The result is consistent with previous analyses of 
community size distributions on large  graphs~\cite{guimera03,arenasrev,palla,clausetfast}, 
although this is the first result concerning the WWW. We stress that we have not performed a complete analysis
of this network, because it would require a lot of processors to carry out the high number of
runs at different $\alpha$-values which are necessary for a reliable analysis. Therefore, the distribution
in Fig.~\ref{fig5} does not necessarily correspond to the most significant cover. However, the $\alpha$-values
of the most representative covers of all networks we have considered turned out to be close to $1$, so the plot
of Fig.~\ref{fig5} is likely to be a fair approximation of the actual distribution.

\section{Conclusions}
\label{sec3}

In this paper we have presented the first method that uncovers simultaneously {\it both} 
the hierarchical and the overlapping community structure of complex networks. 
The method consists in finding the local maxima of a fitness function by 
local, iterative search. The procedure enables each node to be included in more than one 
module, leading to a natural description of overlapping communities. Finally, by tuning 
the resolution parameter $\alpha$ one can probe the network at different scales, exploring
the possible hierarchy of community structure. 
The application of our method to a number 
of constructed and empirical networks has given excellent results.

We would like to emphasize that our method provides a general {\it framework}, 
that yields a large class of algorithms. For instance, one could choose a different 
expression for the fitness function, another criterion to define the most 
meaningful cover, or a different optimization procedure of the fitness for a single cluster. 
The setup we have tested proves to be very reliable, 
but we cannot exclude that different choices yield even better results. In fact, 
the framework is so flexible that it can be easily adapted to the problem at hand: 
if one has hints about the topology of the communities to be found for a specific 
system, this information can be used to design a particular fitness function, 
accounting for the required features of the modules. 

Since the complete analysis of a network's community structure 
can be carried out simultaneously on many computers, the upper size limit  
of tractable graphs can be pushed up considerably.
Our method gives the opportunity to study systematically the distribution of community 
sizes of large networks up to millions of nodes, a crucial aspect of the internal 
organization of a graph, which scholars have just begun to examine. An interesting byproduct of 
our technique is the possibility of quantifying the participation of overlapping nodes in their 
communities by the values of their (node) fitness with respect to each group they belong to. 

Finally, we would like to mention that the method can be naturally extended to 
weighted networks, i.e. networks where links carry a weight. There is no 
need to use any kind of thresholding \cite{Illes}, as the generalization of the 
fitness formula is straightforward: In Eq.~\ref{eq1} we have to replace the degree 
$k$ with the corresponding strength $s$ (expressing the sum over the links' weights). 
Applications to directed 
networks can also be easily devised with suitable choices of the fitness function. Our own function ~\ref{eq1}
could be extended to the directed case, in that one considers the {\it indegree} of the nodes of a subgraph: it is plausible
to assume that the total indegree of the nodes of the subgraph due to links internal to the subgraph exceeds the 
total indegree produced by links coming from external nodes, if the subgraph is a community.

\subsection{Acknowledgments}
We thank Marc Barth\'elemy for 
enlightening discussions and suggestions. We also thank 
A. Arenas, S. G\'omez, A. Pagnani, F. Radicchi and J.J. Ramasco 
for a careful reading of the manuscript. 
JK thanks ISI for hospitality and acknowledges partial support by OTKA K60456.
\begin{appendix}

\section{Dependence on the random seeds}
\label{app1}

The choice of the
random seeds where the community exploration starts 
may affect covers obtained for the same $\alpha$-value. This means in principle that 
we cannot rely on the fitness histogram found for a specific choice of the seeds. We have found that covers 
obtained for different seeds are quite close to each other, and that the most relevant covers that emerge
from the analysis are the same for any choice of the seeds. What may depend on the specific seed adopted
is the ranking of the covers. This can be solved by performing some additional runs of the algorithm for 
different seeds in correspondence to the regions of the $\alpha$-range in which meaningful structures have been
spotted after the first scan. 
The final ranking of the covers is then more reliable than any ranking obtained for a specific choice of the random seeds. 
Since the number of relevant peaks is much smaller than the number of nodes $n$,
the computational cost of the additional runs is negligible as compared to the total number of runs.

\section{Comparing partitions}
\label{app2}

The aim of this section is to discuss the problem of comparing covers.
There are many criteria in the literature (see \cite{melia}), but, to the best of our knowledge, 
the case of overlapping clusters has not been considered yet.
Here, we briefly discuss the issue within the framework of information theory~\cite{mckay}.

The \textit{normalized mutual information} $I_{norm}(X:Y)$~\cite{Danon:2005} is defined as
\begin{equation}
I_{norm}(X:Y)= \frac{H (X) + H(Y) - H(X,Y)}{(H(X)+H(Y))/2}.
\end{equation}
where $H(X)$ ($H(Y)$) is the entropy of the random variable $X$ ($Y$) associated to the partition $\mathcal{C'}$ ($\mathcal{C''}$), whereas
$H(X,Y)$ is the joint entropy. This variable is in the range $[0,1]$ and equals $1$ only when
the two partitions $\mathcal{C'}$ and $\mathcal{C''}$ are exactly coincident.
Another possible similarity measure is the \textit{variation of information}  
$V(X, Y)=H(X|Y)  + H(Y|X)$ \cite{karrer08, melia}. One way to normalize $V(X, Y)$ is
\begin{equation}
\label{norm_v2}
V'_{norm}(X, Y) = \frac{1}{2} \Big( \frac{H(X|Y)}{H(X)} + \frac{H(Y|X)}{H(Y)}  \Big),
\end{equation}
which can be interpreted as the average relative lack of information to infer $X$ 
given $Y$ and vice versa. This normalization will be helpful in the following.

Let us now suppose that a node may belong to more than one cluster. 
The membership of the node $i$ is not a number $\it{x_i} \in \{ $1, 2$  \dots |\mathcal{C'}| \} $ 
anymore, but it must be considered as a binary array of $|\mathcal{C'}|$ entries, 
one for each cluster of the partition $\mathcal{C'}$ (say $({\textbf{x}_i})_k=1$ if the node $i$ is 
present in the $\it{C'_k}$ cluster, $({\textbf{x}_i})_k=0$ otherwise). We can regard the $kth$ entry 
of this array as the realization of a random variable $X_k= (\textbf{X})_k$, whose probability distribution is
\begin{equation}
P(X_k=1)=n_{k} / N \quad P(X_k=0)=1-n_{k} / N,
\end{equation}
where $n_{k}$ is the number of nodes in the cluster $\it{C'_k}$ of $\mathcal{C'}$, i.e. $n_k= |C'_k|$. The same
holds for the random variable $Y_l$ associated to the cluster $\it{C''_l}$ of $\mathcal{C''}$.

It is possible to define the \textit{joint distribution} $P(X_k, Y_l)$
\begin{description}
  \item 
  \begin{equation}
  \label{first_prob}
  P(X_k=1, Y_l=1)= \frac{|C'_k \cap C''_l |}{N},
\end{equation}
  \item 
  \begin{equation}
  P(X_k=1, Y_l=0)= \frac{|C'_k| - |C'_k \cap C''_l |}{N},
\end{equation}
 \item 
  \begin{equation}
  P(X_k=0, Y_l=1)= \frac{|C''_l| - |C'_k \cap C''_l |}{N},
\end{equation}
 \item 
  \begin{equation}
  \label{last_prob}
  P(X_k=0, Y_l=0)= \frac{ N - |C'_k \cup C''_l |}{N}.
\end{equation}
\end{description}
Again, we want to define how similar $\mathcal{C'}$ and $\mathcal{C''}$ are in terms of lack of 
information about one cover given the other. In particular, we can define the amount of 
information to infer $X_k$ given a certain $Y_l$
\begin{equation}
\label{condxkyl}
H(X_k | Y_l)= H(X_k, Y_l) - H(Y_l).
\end{equation}

In order to infer $X_k$, we can choose one $Y_l$ among $|\mathcal{C''}|$ possible candidates. 
In particular, if a cluster $\it{C''_b}$ of $\mathcal{C''}$ turns out to be the same as $\it{C'_k}$, 
we have that $H(X_k | Y_b)=0$, and we would like to say that $Y_b$ is the best candidate to infer $X_k$.  
So, in a set matching fashion, we can decide to consider only $Y_b$ and neglect all the other variables $Y_l$. 
In particular, we can define the conditional entropy of $X_k$ with respect to all the components of $\textbf{Y}$
\begin{equation}
\label{min}
H(X_k | \textbf{Y})= \min_{l  \in \{ 1, 2  \dots |\mathcal{C''}| \} }  H(X_k | Y_l).
\end{equation}
As in Eq. \ref{norm_v2} we can normalize $H(X_k | \textbf{Y})$ dividing by $H(X_k)$
\begin{equation}
H(X_k | \textbf{Y})_{norm}=  \frac{H(X_k | \textbf{Y})}{H(X_k)}
\end{equation}
and taking the average over $k$ eventually leads to the definition of the normalized conditional 
entropy of $\textbf{X}$ with respect to $\textbf{Y}$
\begin{equation}
\label{good_point}
H(\textbf{X} | \textbf{Y})_{norm}= \frac{1}{|\mathcal{C'}|} \sum_k \frac {H(X_k | \textbf{Y})}{H(X_k)}.
\end{equation}
The expression for $H(\textbf{Y} | \textbf{X})_{norm}$ can be determined in the same way. So, we can finally define 
\begin{equation}
\label{final_point}
N(\textbf{X} | \textbf{Y})= 1- \frac{1}{2} [ H(\textbf{X} | \textbf{Y})_{norm} + H(\textbf{Y} | \textbf{X})_{norm}].
\end{equation}
The function $N (\textbf{X} | \textbf{Y}) $ has the appealing property to be equal to one if and only if $X_k=f(Y_l)$ for a certain 
$l$, and vice versa. Unfortunately, this does not imply that $\mathcal{C'}$ and $\mathcal{C''}$ 
are equal. In particular, it may happen that $X_k$ is the \textit{negative} of $Y_l$, i.e. 
\begin{equation}
|C'_k \cap C''_l |= 0 \quad \mbox{and} \quad  |C'_k \cup C''_l |= N.
\end{equation}
In this case we do not need additional information about $X_k$ if we know $Y_l$ because we are sure that if a 
node belongs to $C''_l$ it does not belong to $C'_k$ and vice versa; nevertheless the two covers are not equal.
In other words, taking the minimum in Eq.~\ref{min} may not imply choosing a cluster $C''_b$ very similar to $C'_k$: 
a cluster which is close to the $\textit{complementary}$ of $C'_k$ can be a good candidate as well.

To avoid this problem, we add a constraint in Eq.~\ref{min}: the only eligible $Y_l$ are those ones 
which are far from being the \textit{negatives} of $X_k$, i.e. those fulfilling the following condition
\begin{equation}
\label{constr_h}
h[P(1,1)] +  h[P(0,0)] > h[P(0,1)] +h[P(1,0)],
\end{equation}
where we used the short notation $P(1,1) = P(X_k=1, Y_l=1) \dots$  and $h(p) = -p \log p$. 
To understand why this constraint is appropriate, let us write explicitly the conditional entropy (Eq. \ref{condxkyl})
\begin{eqnarray}
H(X_k|Y_l)&=& h[P(1,1)] +  h[P(0,0)] + h[P(0,1)] +\nonumber\\
&&+h[P(1,0)]  - h[P(Y_l=1)] - h[P(Y_l=0)].
\end{eqnarray}
In the case of $C''_l$ equal to $C'_k$ we have
\begin{equation}
h[P(1,1)] =  h[P(Y_l=1)] \quad \mbox{and} \quad h[P(0,0)] = h[P(Y_l=0)],
\end{equation}
while the mixing terms vanish
\begin{equation}
h[P(0,1)] =0  \quad \mbox{and} \quad h[P(1,0)] =0.
\end{equation}
On the other hand, if $C''_l$ is the complementary to $C'_k$, the role of $h(P(1,1))$ and $h(P(0,0))$ is played by the mixing terms:
\begin{equation}
h[P(0,1)] =  h[P (Y_l=1)] \quad \mbox{and} \quad h[P(1,0)] = h[P (Y_l=0)],
\end{equation}
while
\begin{equation}
h[P(1,1)] =0  \quad \mbox{and} \quad h[P(0,0)] =0.
\end{equation}
So, in the former case all the information quantified by $H(X_k, Y_l)$ is used to encode the \textit{positive} cases 
i.e. $H(X_k, Y_l)= h[P(1,1)] +  h[P(0,0)] $, while in the latter it is used to encode the mixing terms, i.e. 
$H(X_k, Y_l)= h[P(1,0)] +  h[P(0,1)]$. Then, the condition expressed by Eq. \ref{constr_h} means that more than one half 
of $H(X_k, Y_l)$ is used to encode the \textit{positive} cases, and so it excludes the clusters close to being complementary.

If none of the $Y_l$ fulfills Eq. \ref{constr_h}, we set
\begin{equation}
\label{min0}
H(X_k | \textbf{Y})=H(X_k).
\end{equation}
All this assures that $N (\textbf{X} | \textbf{Y}) =1$ if and only if the two covers $\mathcal{C'}$ and $\mathcal{C''}$ are equal.

To sum up, all the procedure reduces to:
\begin{enumerate}
\item for a given $k$, compute $H(X_k|Y_l)$ for each $l$ using the probabilities given by Eqs. \ref{first_prob} $-$ \ref{last_prob};
\item compute $H(X_k | \textbf{Y})$ from Eq. \ref{min} taking into account the constraint given in Eq. \ref{constr_h}; 
note that if this condition is never fulfilled we decided to set $H(X_k | \textbf{Y})=H(X_k)$;
\item for each $k$, repeat the previous step to compute $H(\textbf{X} | \textbf{Y})_{norm}$ according to Eq. \ref{good_point};
\item repeat all this for $\textbf{Y}$ and put everything together in Eq. \ref{final_point}.
\end{enumerate}

\end{appendix}

\end{document}